\documentstyle[12pt]{article}
 1
\font\twelverm=cmr10 scaled\magstep 1
 1

\font\tenbf=cmbx10
\font\tenrm=cmr10
\font\tenit=cmti10

\font\ninerm=cmr9

\font\eightrm=cmr8

\font\teni=cmmi10   \font\seveni=cmmi7  \font\fivei=cmmi5
\font\tensy=cmsy10
\font\sevenbf=cmbx7
\font\fivebf=cmbx5
\font\sevenrm=cmr7

\font\sevensy=cmsy7
\font\fiverm=cmr5

\font\fivesy=cmsy5
\def\tenpoint{\def\rm{\fam0\tenrm}% switch to 10-point type
\textfont0=\tenrm \scriptfont0=\sevenrm \scriptscriptfont0=\fiverm
\textfont1=\teni \scriptfont1=\seveni \scriptscriptfont1=\fivei
\textfont2=\tensy \scriptfont2=\sevensy \scriptscriptfont2=\fivesy
\textfont3=\tenex \scriptfont3=\tenex \scriptscriptfont3=\tenex
\textfont\itfam=\tenit \def\it{\fam\itfam\tenit}
\textfont\slfam=\tensl \def\sl{\fam\slfam\tensl}
\textfont\ttfam=\tentt \def\tt{\fam\ttfam\tentt}
\textfont\bffam=\tenbf \scriptfont\bffam=\sevenbf
\scriptscriptfont\bffam=\fivebf \def\bf{\fam\bffam\tenbf}
\normalbaselineskip=12pt
\setbox\strutbox=\hbox{\vrule height8.5pt depth3.5pt width0pt}
\let\sc=\eightrm \let\big=\tenbig \normalbaselines\rm
}
\catcode`\@=11
\long\def\@makefntext#1{ %\parindent 1em
\protect\noindent \hbox to 3.2pt {\hskip-.9pt
$^{{\ninerm\@thefnmark}}$\hfil}#1\hfill} %can be used

\def\thefootnote{\fnsymbol{footnote}}
 \def\@makefnmark{\hbox to 0pt{$^{\@thefnmark}$\hss}}  %original

\def\ps@myheadings{\let\@mkboth\@gobbletwo
\def\@oddhead{\hbox{} %\sl
\rightmark\hfil\ninerm\thepage}
\def\@oddfoot{}\def\@evenhead{\ninerm\thepage\hfil %\sl
\leftmark\hbox{}}\def\@evenfoot{}
\def\sectionmark##1{}\def\subsectionmark##1{}}

\begin{document}
\newcommand \beq{\begin{eqnarray}}
\newcommand \eeq{\end{eqnarray}}
\newcommand \ga{\raisebox{-.5ex}{$\stackrel{>}{\sim}$}}
\newcommand \la{\raisebox{-.5ex}{$\stackrel{<}{\sim}$}}

%----------------------------PROCSLA.STY---------------------------------------
\newcommand{\symbolfootnote}{\renewcommand{\thefootnote}
	{\fnsymbol{footnote}}}
\renewcommand{\thefootnote}{\fnsymbol{footnote}}
\newcommand{\alphfootnote}
	{\setcounter{footnote}{0}
	 \renewcommand{\thefootnote}{\sevenrm\alph{footnote}}}

%------------------------------------------------------------------------------
%NEW DEFINED SECTION COMMANDS
\newcounter{sectionc}\newcounter{subsectionc}\newcounter{subsubsectionc}
\renewcommand{\section}[1] {\vspace{0.6cm}\addtocounter{sectionc}{1}
\setcounter{subsectionc}{0}\setcounter{subsubsectionc}{0}\noindent
	{\bf\thesectionc. #1}\par\vspace{0.4cm}}
\renewcommand{\subsection}[1] {\vspace{0.6cm}\addtocounter{subsectionc}{1}
	\setcounter{subsubsectionc}{0}\noindent
	{\it\thesectionc.\thesubsectionc. #1}\par\vspace{0.4cm}}
\renewcommand{\subsubsection}[1]
{\vspace{0.6cm}\addtocounter{subsubsectionc}{1}
	\noindent {\rm\thesectionc.\thesubsectionc.\thesubsubsectionc.
	#1}\par\vspace{0.4cm}}
\newcommand{\nonumsection}[1] {\vspace{0.6cm}\noindent{\bf #1}
	\par\vspace{0.4cm}}

%NEW MACRO TO HANDLE APPENDICES
\newcounter{appendixc}
\newcounter{subappendixc}[appendixc]
\newcounter{subsubappendixc}[subappendixc]
\renewcommand{\thesubappendixc}{\Alph{appendixc}.\arabic{subappendixc}}
\renewcommand{\thesubsubappendixc}
	{\Alph{appendixc}.\arabic{subappendixc}.\arabic{subsubappendixc}}

\renewcommand{\appendix}[1] {\vspace{0.6cm}
        \refstepcounter{appendixc}
        \setcounter{figure}{0}
        \setcounter{table}{0}
        \setcounter{equation}{0}
        \renewcommand{\thefigure}{\Alph{appendixc}.\arabic{figure}}
        \renewcommand{\thetable}{\Alph{appendixc}.\arabic{table}}
        \renewcommand{\theappendixc}{\Alph{appendixc}}
        \renewcommand{\theequation}{\Alph{appendixc}.\arabic{equation}}
%       \noindent{\bf Appendix \theappendixc. #1}\par\vspace{0.4cm}}
        \noindent{\bf Appendix \theappendixc #1}\par\vspace{0.4cm}}
\newcommand{\subappendix}[1] {\vspace{0.6cm}
        \refstepcounter{subappendixc}
        \noindent{\bf Appendix \thesubappendixc. #1}\par\vspace{0.4cm}}
\newcommand{\subsubappendix}[1] {\vspace{0.6cm}
        \refstepcounter{subsubappendixc}
        \noindent{\it Appendix \thesubsubappendixc. #1}
	\par\vspace{0.4cm}}

%------------------------------------------------------------------------------
%MACRO FOR ABSTRACT BLOCK  Modified 1/8/94. CLee. \tenrm ==> \tenpoint
\def\abstracts#1{{
	\centering{\begin{minipage}{30pc}\tenpoint\baselineskip=12pt\noindent
	\centerline{\tenpoint ABSTRACT}\vspace{0.3cm}
	\parindent=0pt #1
	\end{minipage} }\par}}

%------------------------------------------------------------------------------
%NEW MACRO FOR BIBLIOGRAPHY
\newcommand{\bibit}{\it}
\newcommand{\bibbf}{\bf}
\renewenvironment{thebibliography}[1]
	{\begin{list}{\arabic{enumi}.}
	{\usecounter{enumi}\setlength{\parsep}{0pt}
%1.25cm IS STRICTLY FOR PROCSLA.TEX ONLY
\setlength{\leftmargin 1.25cm}{\rightmargin 0pt}
%0.52cm IS FOR NEW DATA FILES
%\setlength{\leftmargin 0.52cm}{\rightmargin 0pt}
	 \setlength{\itemsep}{0pt} \settowidth
	{\labelwidth}{#1.}\sloppy}}{\end{list}}

%------------------------------------------------------------------------------
%FOLLOWING THREE COMMANDS ARE FOR 'LIST' COMMAND.
\topsep=0in\parsep=0in\itemsep=0in
\parindent=1.5pc

%LIST ENVIRONMENTS
\newcounter{itemlistc}
\newcounter{romanlistc}
\newcounter{alphlistc}
\newcounter{arabiclistc}
\newenvironment{itemlist}
    	{\setcounter{itemlistc}{0}
	 \begin{list}{$\bullet$}
	{\usecounter{itemlistc}
	 \setlength{\parsep}{0pt}
	 \setlength{\itemsep}{0pt}}}{\end{list}}

\newenvironment{romanlist}
	{\setcounter{romanlistc}{0}
	 \begin{list}{$($\roman{romanlistc}$)$}
	{\usecounter{romanlistc}
	 \setlength{\parsep}{0pt}
	 \setlength{\itemsep}{0pt}}}{\end{list}}

\newenvironment{alphlist}
	{\setcounter{alphlistc}{0}
	 \begin{list}{$($\alph{alphlistc}$)$}
	{\usecounter{alphlistc}
	 \setlength{\parsep}{0pt}
	 \setlength{\itemsep}{0pt}}}{\end{list}}

\newenvironment{arabiclist}
	{\setcounter{arabiclistc}{0}
	 \begin{list}{\arabic{arabiclistc}}
	{\usecounter{arabiclistc}
	 \setlength{\parsep}{0pt}
	 \setlength{\itemsep}{0pt}}}{\end{list}}

%------------------------------------------------------------------------------
%FIGURE CAPTION  Modified 1/8/94. CLee. \tenrm ==> \tenpoint
\newcommand{\fcaption}[1]{
        \refstepcounter{figure}
        \setbox\@tempboxa = \hbox{\tenpoint Fig.~\thefigure. #1}
        \ifdim \wd\@tempboxa > 6in
           {\begin{center}
        \parbox{6in}{\tenpoint\baselineskip=12pt Fig.~\thefigure. #1 }
            \end{center}}
        \else
             {\begin{center}
             {\tenpoint Fig.~\thefigure. #1}
              \end{center}}
        \fi}

%TABLE CAPTION  Modified 1/8/94. CLee. \tenrm ==> \tenpoint
\newcommand{\tcaption}[1]{
        \refstepcounter{table}
        \setbox\@tempboxa = \hbox{\tenpoint Table~\thetable. #1}
        \ifdim \wd\@tempboxa > 6in
           {\begin{center}
        \parbox{6in}{\tenpoint\baselineskip=12pt Table~\thetable. #1 }
            \end{center}}
        \else
             {\begin{center}
             {\tenpoint Table~\thetable. #1}
              \end{center}}
        \fi}

%------------------------------------------------------------------------------
%ACKNOWLEDGEMENT: this portion is from John Hershberger
\def\@citex[#1]#2{\if@filesw\immediate\write\@auxout
	{\string\citation{#2}}\fi
\def\@citea{}\@cite{\@for\@citeb:=#2\do
	{\@citea\def\@citea{,}\@ifundefined
	{b@\@citeb}{{\bf ?}\@warning
	{Citation `\@citeb' on page \thepage \space undefined}}
	{\csname b@\@citeb\endcsname}}}{#1}}

\newif\if@cghi
\def\cite{\@cghitrue\@ifnextchar [{\@tempswatrue
	\@citex}{\@tempswafalse\@citex[]}}
\def\citelow{\@cghifalse\@ifnextchar [{\@tempswatrue
	\@citex}{\@tempswafalse\@citex[]}}
\def\@cite#1#2{{$\null^{#1}$\if@tempswa\typeout
	{IJCGA warning: optional citation argument
	ignored: `#2'} \fi}}
\newcommand{\citeup}{\cite}

%------------------------------------------------------------------------------
%FOR FNSYMBOL FOOTNOTE AND ALPH{FOOTNOTE}
\def\fnm#1{$^{\mbox{\scriptsize #1}}$}
\def\fnt#1#2{\footnotetext{\kern-.3em
	{$^{\mbox{\sevenrm #1}}$}{#2}}}

%----------------------START OF DATA FILE------------------------------

\centerline{\tenbf QUARK MATTER STRUCTURE IN NEUTRON STARS}
\baselineskip=22pt
\vspace{0.4cm}
\centerline{\tenrm H. HEISELBERG \footnote{Talk given at
Intl. Symposium on {\em Strangeness and Quark Matter}, Chania, Greece,
    sept. 1-5 1994, org. A.D.Panagiotou and G.Vassiliadis, proc. to appear in
    World Scientific.} }
\baselineskip=13pt
\centerline{\tenit NORDITA, Blegdamsvej 17, DK-2100 Copenhagen \O, Denmark}
\vspace{0.4cm}
\abstracts{
For physically reasonable bulk and surface properties, quark matter has
recently
been found to coexist with nuclear matter inside neutron stars in a uniform
background of electrons. The microstructure of this mixed
phase starts out with a few
quark matter droplets embedded in the nuclear matter but as the density of
droplets increase towards the center of the neutron star they merge into
rods, plates and other structures. The topology, size as well as Coulomb and
surface energies of these structures  depend sensitively on the quark/nuclear
matter interface tension. A major fraction of the interior of neutron stars
could consist of matter in this new phase, which would provide new mechanisms
for glitches, cooling and supernovae.}
\vglue 0.3cm
%\vspace{0.8cm}
\twelverm   %modified by CLee 23/07/93
\baselineskip=14pt

\section{The Structure of Quark Matter in Neutron Stars}

Over the past two decades many authors\cite{ALL}
have considered the existence
of quark matter in neutron stars. Assuming a first order phase transition one
has, depending on the equation of states, found  either
complete strange quark matter stars or neutron stars with a core of quark
matter
surrounded by a mantle of nuclear matter and a crust on top. Recently, the
possibility of a mixed phase of quark and nuclear matter was considered\cite{G}
and found to be energetically favorable. Including surface and Coulomb energies
this mixed phase was still found to be favored for reasonable bulk and
interface
properties\cite{HPS}. The structure of the mixed phase of  quark matter
embedded
in nuclear matter with a uniform background of electrons was studied and
resembles that in the neutron drip region in the crust. The resulting picture
of
a neutron stars is shown in Fig. 1. Starting
from the outside, the crust consists of the outer layer, which is
a dense solid of neutron rich nuclei, and the inner layer in which
neutrons have dripped and form a neutron gas coexisting with the nuclei.
The structure of the latter mixed phase has recently been calculated in detail
\cite{Rav} and is found to exhibit rod-, plate- and bubble-like structures. At
nuclear saturation density $n_0\simeq 0.16$fm$^{-3}$ there is only one phase of
uniform nuclear matter consisting of mainly neutrons, a small fraction of
protons and
the same amount of electrons to achieve charge neutrality. A mixed phase of
quark matter (QM) and nuclear matter (NM) appear already around a few times
nuclear
saturation density - lower than the phase transition in hybrid stars. In the
beginning only few droplets of quark matter appear but at higher densities
their
number increase and they merge into:  QM rods, QM plates, NM rods,
NM bubbles, and finally pure QM at very high densities if the neutron stars
have not become unstable towards gravitational collapse.

\begin{figure}[thb]

\vspace*{12cm}
\fcaption{The quark and nuclear matter structure in a quarter of a typical
1.4$M_\odot$ solar mass neutron star.
The typical sizes of structures are a few Fermi's but have been scaled
up by about 16 orders of magnitudes to be seen.
\label{star} }
\end{figure}

\section{The Phase Transition}
The phase transition is determined by the  phase coexistence conditions and the
equations of states (EoS) of both NM and QM at essentially zero temperature.
Whether a simple quadratic form\cite{HPS} for the NM EoS or more elaborate EoS
based on Urbana/Argonne potentials, the results only varied quantitatively for
a
wide range of parameters (e.g., compressibility, symmetry energy,  bag
constant,
$\alpha_s$, $m_s$,...). Qualitatively the same picture emerges: {\it The mixed
phase is energetically favored and starts already around twice nuclear
saturation density}.

Here we take for comparison another nuclear EoS, the Walecka relativistic
mean field model\cite{Walecka}. Reproducing the nuclear saturation
density and binding energy determines the scalar
$g_S^2/m_S^2=2300$ MeV/fm$^3$ and vector
$g_V^2/m_V^2=1700$ MeV/fm$^3$ couplings.
%\begin{eqnarray}
%    \epsilon_{NM} &=& .............
%    \epsilon_N =&&n[m+\frac{K_0}{18}(\frac{n}{n_0}-1)^2
% +S_0 (\frac{n}{n_0})^\gamma (1-2x)^2 ] +\frac{\mu_e^4}{12\pi^2}
%       \label{ENM} .
%\end{eqnarray}
%Here $n$ is the baryon density, $n_0=0.16$ fm$^{-3}$ is the nuclear
%saturation density, and $x$ is the proton fraction. The compressibility
%we choose as $K_0\simeq 250$ MeV and for the symmetry term we take that
%of Ref. \cite{Pe91} with $S_0\simeq 30$ MeV and $\gamma\simeq 1$. The
%electron chemical potential is never much above the muon mass and
%therefore muons may be ignored.
For quark matter we assume the bag model equation of state
\begin{equation}
    \epsilon_{QM}=
    (1-\frac{2\alpha_s}{\pi})
    \left(\sum_{q=u,d,s} \frac{3\mu_q^4}{4\pi^2}\right)  + B
    +\frac{\mu_e^4}{12\pi^2}   \, ,
    \label{EQM}
\end{equation}
with the qcd fine structure constant $\alpha_s\simeq 0.4$ and bag
constant $B\simeq 120$ MeV/fm$^3$.
We have taken all quark masses to be zero.

The energy densities of NM and QM are shown in Fig. 2 with two types
of phase transitions depending on the phase coexistence conditions as
will now be discussed.

\subsection{Hybrid Stars}
\vspace*{-3mm}
In hybrid stars it is assumed that each of the two phases are electrically
neutral separately. Thus the proton density is equal to the electron density in
nuclear matter and is given  by $\beta$-equilibrium $\mu_n=\mu_p+\mu_e^{NM}$ at
a given density.  Likewise in the quark matter $\beta$-equilibrium
$\mu_d=\mu_s=\mu_u+\mu_e^{QM}$ and charge neutrality determines all quantities
for a given density.

Gibb's conditions $P_{NM}=P_{QM}$ and $\mu_n^{NM}=\mu_n^{QM}$ (the temperatures
are vanishing in both phases) then determine a unique density at which the two
bulk neutral phases coexist. This is the standard Maxwell   construction and is
seen in Fig. 2 as the double-tangent. In a gravitational field the denser phase
(QM) will sink to the center whereas the lighter phase (NM) will float on top
as
a mantle. At the phase transition there is a sharp density discontinuity and
generally $\mu_e^{NM}\ne\mu_e^{QM}$ so that the electron densities
$n_e=\mu_e^3/3\pi^2$ are {\it different} in the two phases. This assumes that
the sizes of QM structures are larger than electron screening lengths which, as
discussed in \cite{HPS}, is {\it not} the case.

For very low bag constants
the phase transition occur at densities lower than $n_0$ and the
whole neutron star is a quark star except possibly for a hadronic
crust\cite{Weber}.

\subsection{Mixed phase}
\vspace*{-3mm}
We now consider the case where bulk QM is embedded in nuclear matter and the
sizes of the QM structures are smaller than typical electron screening lengths.
Thus the electron background is almost uniform which gives the extra condition
$\mu_e^{NM}=\mu_e^{QM}$. On the other hand charge neutrality is only required
on
average and not in each phase. As a consequence QM and NM can coexist in a wide
range of densities and as seen in Fig. 2(a) the energy density is always lower
than the two bulk neutral phases when surface and Coulomb energies are
neglected
($\sigma = 0$). We also observe that QM in the form of droplets appear at a
lower density than the phase transition between the two bulk neutral phases
with
increasing density or pressure  more and more NM is deconfined into QM.
Defining
the {\it filling fraction } $f=V_{QM}/(V_{NM}+V_{QM})$ as the fraction of the
volume in the QM phase, we see that $f$ increase continuously from zero to
unity
as the density increase from $\sim 2n_0$ to $\sim 9n_0$.

\begin{figure}[thb]

\vspace*{12cm}

\fcaption{(a) The full line gives the energy density of the droplet phase
without surface and Coulomb energies ($\sigma$ = 0). Also shown are the energy
densities of electrically neutral bulk nuclear matter, quark matter in
$\beta$-equilibrium, and  the double tangent construction (dashed line)
corresponding to the coexistence of bulk electrically neutral phases. (b)
Energy
densities of the droplet phase relative to its value for $\sigma$= 0 for
$\sigma$ = 10, 20, and 30 MeV/fm$^2$. When the energy density of the
droplet phase falls within the hatched area it is energetically favored.
\label{phase} } \end{figure}

\section{Why Is the Mixed Phase Energetically Favored?}
This becomes evident by looking at charge densities.
Consider quark matter immersed in a
uniform background of electrons.   Beta equilibrium insures that
$\mu_d=\mu_s=\mu_u+\mu_e$, and therefore in the absence of quark-quark
interactions, one finds the total electric charge density in the quark
matter phase is given for $\mu_e\ll \mu_u\sim\mu_d\equiv\mu_q$ and
$m_s\ll\mu_q$  by
\begin{equation}
 \rho_Q = \frac{e}{3}(2n_u-n_d-n_s-3n_e)
         \simeq \frac{e}{\pi^2}
         \left(\frac{1}{2}m_s^2\mu_q-2\mu_e\mu_q^2\right)
    \,  , \label{rhoq}
\end{equation}
since $n_i=(\mu_i^2-m_i^2)^{3/2}/\pi^2$. Assuming $m_s\simeq 150$ MeV and
$\mu_q\simeq m_N/3$ the second term dominates except for small $\mu_e$ and so
the droplet is negatively charged.
 The electron chemical potential in neutron stars depends
strongly on the model for the nuclear equation of state, but generally one
finds
$\mu_e\raisebox{-.5ex}{$\stackrel{<}{\sim}$} 170$ MeV and so
$\rho_Q\raisebox{-.5ex}{$\stackrel{<}{\sim}$} -0.4e$ fm$^{-3 }$.
Due to the high quark density, $\rho_N$ is small
compared with $\rho_Q$ in Eq. (\ref{R}) when quark matter occupies a small
fraction of the volume.

The QM is therefore {\it negatively} charged and by immersing QM in the
positively charged NM we can either remove some of the electrons from the top
of
the Fermi levels with energy $\mu_e$, or we can increase the proton fraction in
NM by which the symmetry energy is lowered. In equilibrium a combination of
both
will occur and in both cases  {\it bulk energy is saved} and a lower energy
density is achieved as seen in Fig. 2.
However, we still have to pay the Coulomb and surface energies of the
structures, as will now be estimated, and see whether the mixed phase is really
energetically favored.

\section{Properties of the Mixed Phase} \vspace*{-7mm} \subsection{Coulomb and
Surface Energies of Structures} \vspace*{-3mm} Surface and Coulomb energies
determine the topology and length scales of the structures. Denoting the
dimensionality of the structures by $d$ ($d=3$ for droplets and bubbles, $d=2$
for rods and $d=1$ for plates) the surface and Coulomb energies are
generally\cite{Rav}
\beq
    {\cal E}_S&=& d\sigma \frac{4\pi}{3} R^2  \, ,
    \label{ESurf}\\
    {\cal E}_C&=&\frac{8\pi^2}{3(d+2)} (\rho_Q-\rho_N)^2R^5
   \left[\frac{2}{d-2}(1-\frac{d}{2}f^{1-2/d})+f\right] \, ,  \label{ECoul}
\eeq
where $\sigma$ is the surface tension, $R$ the size of the structure, and
$\rho_Q$ and $\rho_N$ are the total charge densities in bulk QM and NM
respectively.  For droplets ($f\simeq0$) or bubbles ($f\simeq1$) $d=3$  and the
Coulomb energies reduce to the usual term ${\cal E}_C=(3/5)Z^2e^2/R$ where $Z$
is the excess charge of the droplet compared with the surrounding medium,
$Ze=(4\pi/3)(\rho_Q-\rho_N)R^3$.
Minimizing the energy density with respect to
$R$ we obtain the usual result that ${\cal E}_S=2{\cal E}_C$\footnote{
The condition for fission instability
is contrarily: $2{\cal E}_S\le {\cal E}_C$ .}.
Minimizing with respect to the continuous dimensionality as well thus
determines both $R$ and $d$.

We now estimate surface and Coulomb energies. When quark matter occupies a
small
fraction of space, $f\simeq 0$,  one can show that the difference in energy
between the droplet phase and bulk neutral nuclear matter varies as $f^2$. In
contrast to this, the contributions to the energy density from surface and
Coulomb energies are linear in $f$. (See Eq. (\ref{energy}))  Similar results
apply for $f$ close to unity.  This shows that the transitions to the droplet
phase must occur via a first-order transition.  However, if the surface and
Coulomb energies are sufficiently large, the droplet phase may never be
favorable. The energy-density difference between the droplet phase, neglecting
surface and Coulomb effects, and two coexisting neutral phases is a few
MeV/fm$^3$, as may be seen from Fig. 2. This is very small compared
with characteristic energy densities which are of order 1000
MeV/fm$^3$.  In Fig. 2(b) we show the energy density of the droplet
phase for various values of the surface tension, relative to the value for
$\sigma = 0$.  For the droplet phase to be favorable, its energy density must
lie below those of nuclear matter, quark matter, and coexisting electrically
neutral phases of nuclear  and quark matter.  That is the droplet phase will be
favored if its energy lies within the hatched region in Fig. 2(a+b).  We see
that whether or  not the droplet phase is energetically favorable depends
crucially on properties of quark matter and nuclear matter.  For our model the
droplet phase is energetically favorable at some densities provided  $\sigma
\raisebox{-.5ex}{$\stackrel{<}{\sim}$} 20$ MeV/fm$^2$.
For comparison, using a quadratic Eos for NM\cite{HPS}
one finds in stead the more favorable condition
$\sigma\raisebox{-.5ex}{$\stackrel{<}{\sim}$} 70$ MeV/fm$^2$.
Given
the large uncertainties in estimates of bulk and surface properties one cannot
at present claim that the droplet phase is definitely favored energetically.

In these analyses several restrictions were made:
the interfaces were sharp, the
charge densities constant in both NM and QM and the background electron
density was also assumed constant.
Relaxing these restrictions generally allow the system to minimize its energy
further.
Constant charge densities may be a
good approximation when screening lengths are much larger than spatial
length scales of structures but since they are only slightly larger\cite{HPS}
the system may save significant energy by rearranging the charges.

\subsection{The Quark and Nuclear Matter Interface Tension}
\vspace*{-3mm}
The surface tension is a crucial but
unfortunately a poorly determined parameter  (see Ref.
\cite{HPS} for a discussion) .  A rough estimate of the surface tension is the
bag constant, $B$, times a typical hadronic length scale $\sim$1 fm
\cite{Young}
and gives $\sigma\simeq 50-450$MeV/fm$^2$. Estimates  from the MIT bag
model and from lattice gauge calculations are somewhat lower. The surface
tension may also  depend on the densities on both sides of the interface.
Whereas the densities inside the neutron star vary from nuclear matter  density
to about an order of magnitude larger, we have checked by detailed computation
with several equation of states that the density difference over the surface
does not vary by much. We therefore keep the surface tension as an unknown but
density independent parameter.

\subsection{Droplet Radii, Charge and Mass Numbers}
\vspace*{-3mm}
When $f\sim 0$ spherical QM droplets form in NM
whereas when $f\sim 1$ bubbles of NM are embedded in QM.
The surface energy and Coulomb energies
are given by Eqs. (\ref{ESurf}) and (\ref{ECoul}) for $f=0$ or $f=1$ and
the minimization condition ${\cal E}_S=2{\cal E}_C$ gives
\begin{eqnarray}
    R&=&\left(\frac{15}{8\pi}\frac{\sigma}{
       (\rho_Q-\rho_N)^2}\right)^{1/3}
      \simeq 5.0\, {\rm fm} \, \left(\frac{\sigma}{\sigma_0}\right)^{1/3}
      \left( \frac{\rho_Q-\rho_N}{\rho_0}\right) ^{-2/3}.    \label{R}
\end{eqnarray}
In the second formula we have introduced the typical
quantities $\rho_0=-e~ 0.4\cdot$
fm$^{-3}$ and $\sigma_0 = 50$ MeV/fm$^2$. (Symmetric
nuclear matter in vacuum has a surface tension $\sigma=1$
MeV/fm$^2$ for which (\ref{R}) gives $R\simeq$ 4 fm which
agrees with the fact that nuclei like $^{56}Fe$ are the most stable.)
The total Coulomb and surface energy per unit volume
is given for small $f$ by
\begin{eqnarray}
     \epsilon_{S+C}&=& f\, 9\left(\frac{\pi}{15}\sigma^2
(\rho_Q-\rho_N)^2\right)^{1/3}
\, \simeq 44\, {\rm MeV~fm^{-3}} \, f \,
      \left(\frac{\sigma}{\sigma_0}
 \frac{\rho_Q-\rho_N}{\rho_0}\right) ^{2/3}.    \label{energy}
\end{eqnarray}
The result for $f$ close to unity is given by replacing $f$ by $1-f$.
The case when the volumes of quark and nuclear matter
are equal, i.e. alternating layers of QM and NM ($f=1/2$),
is considered in \cite{HPS}. Similar length scales but smaller Coulomb
and surface energies are found.

Consequently, for $\sigma \simeq 10$ MeV/fm$^2$ we find from Eq.
(\ref{R}) a radius of $R\raisebox{-.5ex}{$\stackrel{>}{\sim}$}$ 3.1 fm, whereas
$\sigma \simeq 100$ MeV/fm$^2$ gives
$R\raisebox{-.5ex}{$\stackrel{>}{\sim}$}$ 6.6 fm.  As $\mu_e$ decreases with
increasing density the length scales increase as $\propto\mu_e^{-2/3}$. The
corresponding mass number and charge are a few hundreds and up.

\subsection{Melting Temperature}
\vspace*{-3mm}
The melting temperature of a bcc Coulomb lattice is\cite{melt}
\beq
   T_c \simeq\frac{Z^2e^2}{170a} = \frac{Z^2e^2}{170R}f^{1/3} \, , \label{Tc}
\eeq
where $a$ is the distance between lattice points. The large numerical factor of
170 reflects the fact that it only takes a fraction of the usual Coulomb energy
$Z^2e^2/R$ for two atoms to slide by each other in a lattice. $T_c$ is
typically
some hundreds of MeV - much larger than temperatures inside neutron stars,
which are estimated to reach $\sim$10MeV in supernovae  cool rapidly
thereafter.
The mixed phase is therefore {\it frozen solid}. Already at densities $\sim
2n_0$ a lattice of QM droplets form  which is only melted for very small $f$ or
equivalently long lattice distance. However, when the Debye screening length of
electrons or protons\cite{HPS} is shorter than the lattice distance the droplet
charge is effectively screened off and a neutral object is formed, which may
diffuse around in all directions.

\section{Summary and Consequences} Assuming a first order phase transition a
mixed phase of quark and nuclear matter  is energetically favored for a wide
range of equation of states - unless the surface tension is too large,
$\sigma\ga 70$MeV/fm$^2$.
Quantitative calculations depends on the equation of states applied but
typically
already around a few times nuclear saturation density
droplets of quark matter appear in a lattice embedded in nuclear matter in a
uniform background of electrons. The existence of such a mixed phase may have a
number of consequences:

\vspace*{-3mm}
\subsection{Glitches}
\vspace*{-3mm}
The solidity of the mixed phase affects quake phenomena, which have been
invoked
to explain observations in pulsar glitches. Some features have been explained
by a simple two-component model of a rotating neutron star that gradually slows
down\cite{Baym} and becomes less deformed which strains the rigid component
(original believed to be the lattice in the crust).
In this model the lattice suddenly cracks/quakes and changes its structure
towards being more spherical. Consequently, its moment of inertia, $I_c$, is
decreased
and its rotation and pulsar frequency increased due to angular momentum
conservation.  Subsequently, the two components slowly relaxates
to a common rotational frequency on a timescale of days due to
superfluidity of the other component (the neutron liquid).
The {\it healing parameter} $Q=I_c/I_{tot}$
measured in glitches reveals that for the Vela and Crab pulsar about $\sim$3\%
and $\sim$96\%  of the moment of inertia is in the rigid component
respectively.
Previously the crust was assumed to be the only rigid component and so the Vela
should be almost all crust. This would require that the Vela is a very light
neutron star - much smaller than the observed ones which all are compatible
with
$\sim 1.4M_\odot$.
If we by the lattice component include not only the the solid crust but
also the protons in NM
(which is locked to the crust due to magnetic fields) and the solid
QM mixed phase
\beq
   I_c = I_{crust}+I_p+I_{QM} \, ,
\eeq
we can better explain the large $I_c$ for the Crab.
The moment of inertia of the mixed phase is sensitive to the EoS's used.
For example, for a quadratic NM EoS\cite{HPS}
increasing the Bag constant from 95 to 110MeV/fm$^3$ reduces
$I_c/I_{total}$ from $\sim70\%$ to $\sim20\%$
for a 1.4$M_\odot$ neutron star.
The structures in the mixed phase would exhibit anisotropic elastic properties,
being rigid to some shear strains but not others in much the same way as liquid
crystals. Therefore the whole mixed phase might not be rigid.

Furthermore, the energy released in glitches every few years
are too large to be stored in the crust only.
The recurrence time for giant quakes, $t_c$, is inversely
proportional to the strain energy\cite{Pines},
which again is proportional to the lattice density and the Coulomb energy
\beq
   t_c^{-1} \propto  \frac{1}{a^3} \frac{Z^2e^2}{a} \, .
\eeq
Since the lattice distance is smaller for the quark matter droplets
and their charge larger
than for atoms in the crust, the recurrence
time is shorter in better agreement with measurements of giant glitches.

So far this is all just speculation and other models as e.g. superfluid
vortices
pinned to the crust\cite{AI} have been invoked to explain glitches.
Detecting core and crust quakes separately or other
signs of three components in glitches, indicating the existence of a
crust, superfluid neutrons and a  solid core, would support the idea of the
mixed quark and nuclear matter mixed phase. However, magnetic field attenuation
is expected to be small in neutron stars and therefore magnetic fields
penetrate
through the core. Thus the crust and core lattices as well as the proton liquid
should be strongly coupled and glitch simultaneosly.

\subsection{Cooling}
\vspace*{-3mm}
Neutrino generation, and hence cooling of neutron stars could be influenced by
the mixed phase. This could come about because nuclear matter in the droplet
phase has a higher proton concentration than bulk neutral nuclear matter and
this could make it easier to attain the threshold condition for the nucleon
direct Urca process\cite{Pe91}. Another is that the presence of the spatial
structure of the droplet phase might allow processes to occur which would be
forbidden in a translationally invariant system. Also the mere presence
of quark matter can lead to fast cooling\cite{qcool} when $\alpha_s\ne0$.
All these mechanisms lead to faster cooling.

\subsection{Maximum mass and Rotational Speed of Neutron Stars}
\vspace*{-3mm}

The EoS is softened by the phase transition to QM
which in both
strange stars\cite{Weber} and hybrid stars\cite{ALL}
leads to lighter maximum mass neutron stars.
The mixed phase has an even softer EoS as that of the double tangent
construction in hybrid stars and has therefore a slighly lighter maximum
mass\cite{Vijay}. In all cases, however, the maximum mass depends
strongly on the EoS of nuclear and quark matter.

The maximum rotation rate and damping of radial density oscillations\cite{osc}
depend on bulk and shear viscosities. These in turn depend on the structures
inside the mixed phase. As discussed above the viscosities can be enormous for
a rigid lattice but might entirely vanish in plate-like structures
that may behave as a liquid crystal.

\subsection{Supernovae}
\vspace*{-3mm}
In a supernova the core collapse is stopped by the incompressibility of nuclear
matter. The softer the equation of state the denser the matter is compressed
before it bounces and the deeper into the gravitational well the star has
fallen. Also a softer EoS creates a more coherent shock wave that excites the
matter less. Consequently, more gravitational energy is available and can
be transferred to neutrino generation which is believed to power the
supernova explosion. Besides softening the EoS it was mentioned above that the
mixed phase occurred through a first order phase transition. Thus latent heat
can be stored which may also affect supernovae.

Due to neutrino trapping and  non-zero temperatures  the situation is, however,
somewhat different in supernovae core collapse than in old neutron stars.  In
particular the high density of neutrinos in  $\beta$-equilibrium increase the
energy densities and pressures so that typically only around twice nuclear
matter densities are reached in cores of supernovae. Thus the amount of quark
matter in the newly formed  neutron star will be less if any, and the supernova
explosion will be less affected as well.

\section{Acknowledgements}
I thank Espen Staubo and Chris Pethick who collaborated on much on the work
reported here.
This work was supported in part by the Director, Office of Energy Research,
Office of High Energy and Nuclear Physics, Division of Nuclear Physics,
of the U.S. Department of Energy under Contract DE-AC03-76SF00098,
and the Danish Natural Science Research Council.

\end{document}